\documentclass[twocolumn]{aastex6}

\usepackage{graphicx,amsmath,color}
\usepackage[caption=false]{subfig}
\usepackage{amssymb}

\begin{document}

\title[T~Tau at 149\,MHz]{A LOFAR detection of the low mass young star T~Tau at 149\,MHz}
\author{Colm P. Coughlan\altaffilmark{1}, Rachael E. Ainsworth\altaffilmark{1}, Jochen Eisl{\"o}ffel\altaffilmark{2}, Matthias Hoeft\altaffilmark{2}, Alexander Drabent\altaffilmark{2}, Anna M. M. Scaife\altaffilmark{3}, Tom P. Ray\altaffilmark{1}, Martin E. Bell\altaffilmark{4,5}, Jess W. Broderick\altaffilmark{6}, St{\'e}phane Corbel\altaffilmark{7,8}, Jean-Mathias Grie{\ss}meier\altaffilmark{9,8}, Alexander J. van der Horst\altaffilmark{10}, Joeri van Leeuwen\altaffilmark{6,11}, Sera Markoff \altaffilmark{11}, Malgorzata Pietka\altaffilmark{12}, Adam J. Stewart\altaffilmark{12}, Ralph A.M.J. Wijers\altaffilmark{11} and Philippe Zarka\altaffilmark{8,13}}

\affil{$^1$Dublin Institute for Advanced Studies, School of Cosmic Physics, 31 Fitzwilliam Place, Dublin D02 XF86, Ireland \\
$^2$Th{\"u}ringer Landessternwarte, Sternwarte 5, 07778, Tautenburg, Germany \\
$^3$Jodrell Bank Centre for Astrophysics, School of Physics and Astronomy, The University of Manchester, Oxford Road, Manchester M13 9PL, UK\\
$^4$CSIRO Astronomy and Space Science, PO Box 76, Epping, NSW 1710, Australia\\
$^5$ARC Centre of Excellence for All-sky Astrophysics (CAASTRO), The University of Sydney, NSW 2006, Australia\\
$^6$ASTRON, the Netherlands Institute for Radio Astronomy, Postbus 2, 7990 AA Dwingeloo, The Netherlands\\
$^7$Laboratoire AIM (CEA/IRFU - CNRS/INSU - Universit\'e Paris Diderot), CEA DSM/IRFU/SAp, F-91191 Gif-sur-Yvette, France\\
$^8$Station de Radioastronomie de Nan\c{c}ay, Observatoire de Paris, PSL Research University, CNRS, Univ. Orl\'{e}ans, 18330 Nan\c{c}ay, France\\
$^9$LPC2E - Universit\'{e} d'Orl\'{e}ans / CNRS, 45071 Orl\'{e}ans cedex 2, France\\
$^{10}$Department of Physics, The George Washington University, 725 21st Street NW, Washington, DC 20052, USA\\
$^{11}$Anton Pannekoek Institute for Astronomy, University of Amsterdam, Postbus 94249, 1090 GE Amsterdam, The Netherlands\\
$^{12}$Astrophysics, Department of Physics, University of Oxford, Keble Road, Oxford OX1 3RH, UK\\
$^{13}$LESIA, Observatoire de Paris, CNRS, PSL/SU/UPMC/UPD/SPC, Place J. Janssen, 92195 Meudon, France}

\begin{abstract}
Radio observations of young stellar objects (YSOs) enable the study of ionised plasma outflows from young protostars via their free--free radiation. Previous studies of the low-mass young system T~Tau have used radio observations to model the spectrum and estimate important physical properties of the associated ionised plasma (local electron density, ionised gas content and emission measure). However, without an indication of the low-frequency turnover in the free--free spectrum, these properties remain difficult to constrain. This paper presents the detection of T~Tau at 149\,MHz with the Low Frequency Array (LOFAR) - the first time a YSO has been observed at such low frequencies. The recovered total flux indicates that the free--free spectrum may be turning over near 149\,MHz. The spectral energy distribution is fitted and yields improved constraints on local electron density ($(7.2 \pm 2.1)\times10^{3}$\,cm$^{-3}$), ionised gas mass ($(1.0 \pm 1.8)\times10^{-6}$\,M$_{\odot}$) and emission measure ($(1.67 \pm 0.14)\times10^5$\,pc\,cm$^{-6}$).
\end{abstract}

\maketitle

\section{Introduction}
\label{sec:introduction}

Classical T Tauri stars are pre-main-sequence stars which grow in mass through accretion from a surrounding circumstellar disk. They are observed to drive jets which are believed to remove excess angular momentum from the circumstellar disk thereby allowing accretion to proceed. These Young Stellar Objects (YSOs) are typically detected at centimetre or longer wavelengths via free--free emission from their collimated, ionised outflows \citep[see e.g.][]{Anglada2015}. The observed free--free spectrum is characterised by a flat or positive power law spectral index $\alpha$ (where the flux density $S_{\nu}\propto\nu^{\alpha}$ at frequency $\nu$) with a value of $-0.1$ for optically thin emission, $+2$ for optically thick emission and intermediate values for partially opaque plasma \citep[see e.g.][]{Scaife2013}. 

Historically, with typical flux densities at centimetre wavelengths of $\la1$\,mJy and a tendency to weaken at lower frequencies, observations of YSOs have been confined to frequencies $>1$\,GHz (wavelengths $\lambda<30$\,cm). However, with their free--free spectra expected to turnover at low frequencies as the emitting medium transitions from optically thin to thick \citep{Scaife2013}, the study of YSOs at low frequencies offers a promising window into the physical parameters of their jets. Recent work by \citet{Ainsworth2014, Ainsworth2016} has used the Giant Metrewave Radio Telescope (GMRT) to begin studying YSOs at very low radio frequencies (323 to 608\,MHz) and shows the feasibility of observing YSOs with interferometers such as the Low Frequency Array (LOFAR) and, ultimately, the Square Kilometre Array (SKA).

\citet{Ainsworth2016} discussed the importance of investigating the radio emission from young stars at low radio frequencies. Physical properties of the ionised plasma, such as the electron density, ionised gas mass and emission measure, can be well constrained once the low-frequency turnover (between optically thin and thick behaviour) in the free--free spectrum has been identified. These GMRT results demonstrated that the low frequency turnover occurs at frequencies lower than 323\,MHz for their small target sample. Specifically, these authors showed that the prototype of an entire class of low-mass YSOs \citep{Joy1945}, T~Tau, would be well suited for further investigation with LOFAR as they estimate the turnover frequency to occur around $\sim200$\,MHz. They identified the sensitivity and resolution of LOFAR at 150\,MHz as being well suited to constrain the size of the free--free emitting region in T~Tau.

Low frequency observations of young stars are also important for another reason: searching for non-thermal (synchrotron) emission from YSO jets \citep[see e.g.][]{Carrasco-Gonzalez2010}. This result was originally considered surprising, since the velocities of YSO jets are much lower than those from Active Galactic Nuclei, however recent modelling \citep[e.g.][]{Padovani2016} have shown how YSO jets can accelerate particles to relativistic energies through diffusive shock acceleration. Moreover, the successful detection of polarised emission associated with non-thermal processes is now considered an important window into the magnetic field environment of the jet. It is also notable that a large fraction of YSOs detected with the Very Large Array (VLA) in the Gould Belt Survey of \citet{Dzib2014} are consistent with non-thermal coronal emission. Flux densities from non-thermal emission processes increase with decreasing frequency which should make them easier to detect with instruments such as LOFAR and the GMRT. 

T~Tau (J2000: 04\,21\,59.4 +19\,32\,06.4) is a well-studied triple system located in the Taurus-Auriga Molecular Cloud at a distance of 148\,pc \citep{Loinard2007b}. The triplet consists of an optically visible star, T~Tau~N, and an optically obscured binary $0\farcs7$ ($\approx100$\,au) to the south, T~Tau~Sa and Sb, visible at infrared and longer wavelengths and with a projected separation of $0\farcs13$ \citep[$\approx19$\,au;][]{Dyck1982, Koresko2000, Kohler2008}. Two large-scale outflows have been observed from the system \citep{Reipurth1997, Gustafsson2010}: an east-west outflow from T~Tau~N which terminates at the Herbig-Haro object HH~155 (position angle $\sim 65^{\circ}$ with inclination angle to the plane of the sky $i\sim80^{\circ}$), and a southeast-northwest flow associated with HH~255 and HH~355 thought to emanate from T~Tau~Sa (position angle $\sim 345^{\circ}$ with $i\leq10^{\circ}$). A separate molecular flow in the southwest direction is associated with T~Tau~Sb \citep{Gustafsson2010}. T~Tau~N appears to be surrounded by an accretion disk that is nearly face-on, T~Tau~Sa is surrounded by a disk that is nearly edge-on, and both T~Tau~Sa and T~Tau~Sb are apparently surrounded by an edge-on circumbinary torus which is responsible for the large extinction toward T~Tau~S \citep[see e.g.][for an overview of the T~Tau system]{Loinard2007a}.

\begin{figure}
\begin{center}
\includegraphics[width=\columnwidth]{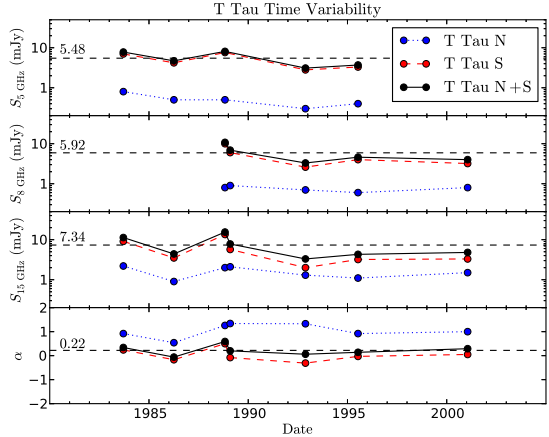}
\caption{Time dependence of the flux density (at 5, 8 and 15\,GHz) and spectral index ($\alpha$, calculated between 5 and 15\,GHz except for epochs 1989 and 2001 which were calculated between 8 and 15\,GHz) of the T~Tau system from \citet[][data are from their Table~2]{Johnston2003}. The (blue) dotted lines show the measurements for T~Tau~N, the (red) dashed lines show the measurements for T~Tau~S and the (black) solid lines shows the combined, total measurement of the (T~Tau~N+S) system. The mean values for the flux densities and spectral index of the total T~Tau~N+S system are quoted on the left of each plot and shown as horizontal (black) dashed lines. It is clear that the total flux density and spectral index are dominated by the variable emission from T~Tau~S (which is more extreme at higher frequencies) and that any unresolved observations are therefore expected to be heavily dominated by T~Tau~S in a ratio of approximately 6:1 (see text for details).}
\label{fig:ttau_var}
\end{center}
\end{figure}

At radio (centimetre) wavelengths, previous high angular resolution observations made with the VLA resolve the northern and southern components of the T~Tau triple system but do not resolve the southern binary \citep{Schwartz1986, Skinner1994, Johnston2003, Smith2003, Loinard2007a}. Figure~\ref{fig:ttau_var} presents the multi-epoch observations at 5, 8 and 15\,GHz of \citet{Johnston2003}. The emission from T~Tau~N does not exhibit a large degree of time variability and has a spectral index $\alpha\simeq1$, indicating that the emission is thermal in origin \citep[e.g. from a dense ionised stellar wind, see also][]{Loinard2007a}. In contrast, the emission from T~Tau~S shows a high degree of time variability over $\simeq20$\,years and usually has a flat spectral index ($\alpha\simeq0$) which is normally indicative of optically thin free--free radiation. However, both T~Tau~N and S radio sources exhibit circular polarisation indicating that the emission from each component has contributions from gyrosynchrotron radiation due to magnetic activity \citep{Skinner1994, Ray1997, Johnston2003}. 

At all epochs, the southern radio source is heavily dominant (with flux density ratio of order 6:1 between the southern and northern radio components, see Figure~\ref{fig:ttau_var}) and therefore makes the major contribution to the spectrum (and the variability) of the unresolved observations \citep[lower resolution observations suggest a ratio 10:1, see e.g.][]{Scaife2011}. \citet{Johnston2003} suggest that the southern radio source itself is dominated by the optically less luminous member, T~Tau~Sb. This is further supported by very high angular resolution observations using the Multi-Element Radio-Linked Interferometer Network (MERLIN) and the Very Long Baseline Array (VLBA) which trace the proper motion of the southern radio source and shows it to be associated, although not coincident, with T~Tau~Sb \citep{Smith2003, Johnston2004, Loinard2005, Loinard2007b}. These authors show that this small scale ($\lesssim1$\,au), compact emission arises from non-thermal gyrosynchrotron radiation implying a magnetic origin. However, there is another radio component which is extended (with position angle of $47^\circ$ at 8\,GHz), unpolarised, only moderately variable and has a spectral index typical of optically thin free--free radiation \citep[$\alpha\simeq0$;][]{Loinard2007a}. This extended, large-scale component is presumed to be diffuse optically thin emission produced by shocks in the supersonic outflow driven by T~Tau~Sb. Based on these results, we would therefore expect large-scale radio emission from the (unresolved) T~Tau system to be dominated by the shock ionised plasma of the T~Tau~Sb outflow, with only small contributions from the compact, non-thermal component of T~Tau~Sb and the thermal component from T~Tau~N.

The recent GMRT observations of T~Tau at 323 and 608\,MHz by \citet{Ainsworth2016} did not resolve the separate components of the YSO, but detected an extended source with a 323--608\,MHz spectral index of $\alpha_{\rm GMRT}=0.11\pm 0.14$, consistent with (partially) optically thin free--free emission. These results are consistent with the extended free--free component of the emission from the T~Tau~Sb outflow \citep{Loinard2007a} which is expected to dominate on large scales (several hundred au). Based on their modelling from 323\,MHz to $\sim1$\,THz, \citet{Ainsworth2016} estimated that the frequency of the turnover in the free--free spectrum of this emission is $\sim200$\,MHz, suggesting LOFAR observations may help constrain the turnover frequency and greatly improve spectral modelling.

In this paper we present the first detection of a YSO with LOFAR at 149\,MHz. In Section~\ref{sec:observations} we present details of the observations and data reduction, as well as a discussion on the accuracy of the recovered LOFAR fluxes. In Section~\ref{sec:results} we present the resulting radio image of the T~Tau system made at 149\,MHz (2\,m) with a bandwidth of 49\,MHz, along with a measurement of the integrated flux density. In Section~\ref{sec:discussion} we model the spectral energy distribution (SED) and perform an analysis of the fitting parameters. As a result, we place limits on the average electron density, the mass of ionised gas and the emission measure, using a fitted turnover frequency of $157 \pm 27$\,MHz. We make our concluding remarks in Section~\ref{sec:conclusions}.

\section{Observations and data reduction}
\label{sec:observations}

The T~Tau system was observed for 8 hours with LOFAR \citep{VanHaarlem2013} in high band (HBA) mode over a single night on 2014 January 9--10 (Project code LC1\_001). 3C147 was observed as a flux calibrator at the beginning and end of the run. A single beam was used to observe the target field between 110 and 190\,MHz using 488 sub-bands with 1 second integration and 64 channels per sub-band. The data were processed by the default pre-processing pipeline which performed flagging using \textsc{aoflagger} \citep{Offringa2012} and averaged the data down to 5 seconds integration and 4 channels per sub-band. While international stations were included in the observation, only data from the Dutch (core and remote) LOFAR stations are used in this paper. This results in a resolution which can be directly compared with the GMRT results of \citet{Ainsworth2016}.

Amplitude and clock calibration were performed using the pre-facet calibration pipeline \textsc{prefactor} \citep{Weeren2016}. This pipeline used the bright calibrator source 3C147 (observed in two 10 minute scans) to calculate the amplitude gains, as well as to separate the clock and TEC (total electron content) phase delay terms from the overall phase delay. As the quality of the second (later) scan was much better than the first, only this scan was used to derive the amplitude and clock solutions to be applied to the target field. Note that the TEC delay solutions were not transferred as they are not applicable to the target field. The diagnostic plots produced by \textsc{prefactor} for the lowest and highest 10\,MHz of the bandwidth suggested they were of poor quality and they were excluded from any further processing.

An initial direction-independent phase calibration was applied to the target field using data from the LOFAR global sky model \citep[GSM;][]{VanHaarlem2013}, a fusion of the VLA Low-frequency Sky Survey \citep[VLSS, VLSSr;][]{Cohen2007,Lane2012}, the Westerbork Northern Sky Survey \citep[WENSS;][]{Rengelink1997} and the NRAO VLA Sky Survey \citep[NVSS;][]{Condon1998}. A sky model at LOFAR frequencies was then developed using the data itself. The field was imaged with \textsc{CASA} \citep{McMullin2007} at five 2\,MHz bands at regular intervals across the most sensitive region of bandwidth (with central frequencies of 127.6, 139.4, 151.1, 162.8 and 172.6\,MHz, respectively) and the LOFAR source finding software, \textsc{PyBDSM} \citep{Mohan2015}, was used to make a multi-frequency sky model of the field. The resulting sky model was then used to generate a new set of direction-independent phase solutions for the original dataset and the process was iterated until two rounds of direction-independent phase-only self-calibration had been applied to the dataset. In all cases the standard corrections for the LOFAR beam were applied immediately before imaging.

Several of the brightest sources with strong artefacts were then subtracted using \textsc{SAGECal} \citep{Kazemi2011} with the sky model used in the final direction-independent self-calibration cycle. To achieve the best subtraction and account for a known issue in direction-dependent calibration where the flux of unmodelled sources can be suppressed, \textsc{SAGECal} was run in ``robust'' multi-frequency Message Passing Interface (MPI) mode \citep{Kazemi2013}, minimising flux suppression and ensuring smooth solutions between neighbouring bands. A bright off-field source, 3C123, was also subtracted. Sources to be subtracted were divided into individual \textsc{SAGECal} solution clusters with unique amplitude and phase solutions, while a single cluster with a single set of amplitude and phase solutions was used to describe the rest of the field. An integration time of 20 minutes was employed and the solutions were applied to the visibilities of the bright sources before subtracting them. The derived solutions for the rest of the field were not applied to the residual visibilities so as to minimise any effect on the target itself. 

Final imaging was performed using \textsc{CASA} over approximately 49\,MHz of bandwidth centred at 149.1\,MHz (sub-bands 50-300). Only core and remote stations were used and the UV range was clipped to above 1\,k$\lambda$. This increased the effective resolution of the image, while suppressing diffuse emission on large scales relative to the anticipated size of the T~Tau system in the GMRT images of \citet{Ainsworth2016}. Three Taylor terms were employed to describe the sky brightness and 256 W-projection planes were used to image $2^{\circ}$ of the sky with a cellsize of $1\arcsec$. The data were weighted with a Briggs Robustness value of zero.

Note that \textsc{AWImager} \citep{Tasse2013}, a LOFAR imager which can apply further corrections for the LOFAR beam during imaging, was also used to image T~Tau. However, as \textsc{AWImager} does not support multifrequency synthesis imaging with two or more Taylor terms, this resulted in reduced quality due to the large fractional bandwidth in the image. As the \textsc{CASA} image appeared generally of higher quality and primary beam effects at the phase centre tend to be small, the result from \textsc{CASA}'s \textsc{clean} task using multi-frequency synthesis was deemed to be the most accurate. \textsc{PyBDSM} was used to fit the emission from the target, resulting in the Gaussian parameters reported in Section~\ref{sec:results}.

\begin{figure}
\begin{center}
\includegraphics[width=\columnwidth]{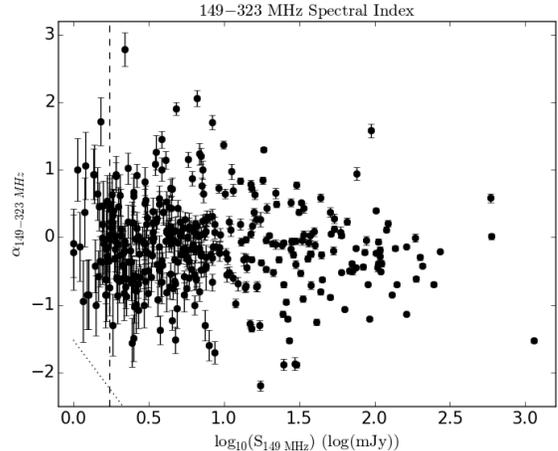}
\caption{Distribution of spectral index vs. flux density at 149\,MHz for all sources detected at 149\,MHz with LOFAR and in the 323\,MHz GMRT image of \citet{Ainsworth2016}. The dashed line indicates the position of T~Tau with an integrated flux of $1.75 \pm 0.36$\,mJy. The dotted line shows the $3\,\sigma$ sensitivity cutoff for the 323\,MHz GMRT image ($\sigma_{\rm rms} = 103\,\mu$Jy\,beam$^{-1}$). Error bars indicate $1\,\sigma$ non-systematic errors propagated from fits by \textsc{PyBDSM}. There is no evidence for any systematic variation of spectral index with flux, suggesting that significant faint sources in the LOFAR image are free of systematic errors relative to the brighter sources.}
\label{fig:lofar-gmrt}
\end{center}
\end{figure}

CASA was used to measure the standard deviation of the flux in a $144$\,arcmin$^{2}$ box around the T~Tau system, finding a value of $\sigma_{\rm rms} = 0.2$\,mJy\,beam$^{-1}$. Figure~\ref{fig:lofar-gmrt} shows the spectral indices derived for sources detected in both the two degree LOFAR image and the 323\,MHz GMRT image of \citet{Ainsworth2016}. No systematic change in spectral index is evident as sources approach the $3\,\sigma_{\rm rms} = 0.6$\,mJy\,beam$^{-1}$ detection limit of the 149\,MHz LOFAR image. This demonstrates that, even without primary beam correction, there is no visible artificial suppression or enhancement of the fluxes of weak sources such as T~Tau, at least within the central part of the image.

While Figure \ref{fig:lofar-gmrt} shows a stable flux scale across a wide range of fluxes, the absolute accuracy of the local flux scale still needs to be established. Two measures were taken to test this accuracy. First; in order to quantify errors due to elevation-dependent gain, the clock and gain solutions were applied to the previously unused first scan of 3C147 (taken 8 hours before the second calibrator scan). Phase only self-calibration and LOFAR beam corrections were applied in the same way as for the T~Tau field. 3C147 was imaged using CASA, recovering a flux of $58.8$\,Jy at 149.1\,MHz compared to the calibrator model flux of $66.7$\,Jy at 150\,MHz, estimated to be accurate to within 4\% of the true flux \citep{Scaife2012a}. The mean elevations of the first and second 3C147 scans were $\sim 40^{\circ}$ and $\sim 64^{\circ}$, respectively, compared with T~Tau's mean elevation of $\sim 45^{\circ}$. This implies that the elevation-dependent gain error on the target field is likely less than the 12\% observed between the two 3C147 scans. No correction was applied to the target field based on this analysis, but the factor of 12\% was considered to be contributing to the systematic uncertainty (see below).

Second; in order to quantify imperfect modelling of LOFAR beam variations across direction and frequency, the integrated LOFAR fluxes of sources within a degree of T~Tau at 149\,MHz were compared to the GMRT 150\,MHz TGSSADR survey, which has an estimated flux scale uncertainty of 10\% \citep{Intema2016} and which also uses the flux scale of \citet{Scaife2012a}. 15 matches were detected within $0\fdg5$ of T~Tau, with LOFAR overestimating the GMRT flux by an average of 2\%. 53 matches were detected within $1^{\circ}$ (inclusive of the 15 within $0\fdg5$), with LOFAR underestimating the GMRT flux by an average of 6\%. Given this good agreement with TGSSADR 150\,MHz fluxes, the LOFAR flux scale is likely accurate to within about 10\%. We thus conservatively estimate the LOFAR flux scale to be accurate to 12\% in agreement with the 3C147-based testing and have added this figure in quadrature to the Gaussian fit uncertainty, $\sigma_{\rm fit}$, to calculate the final uncertainty in the integrated flux: $\sigma_{S_{\nu}} = \sqrt{\sigma_{\rm fit}^{2} + (0.12 \times S_{\rm \nu,int})^{2}}$. Note that $\sigma_{\rm fit}$ is derived from both the quality of the Gaussian fit and the local root-mean-squared noise as determined by \textsc{PyBDSM}.

\section{Results}
\label{sec:results}

\begin{figure*}
\begin{center}
\includegraphics[width=\textwidth]{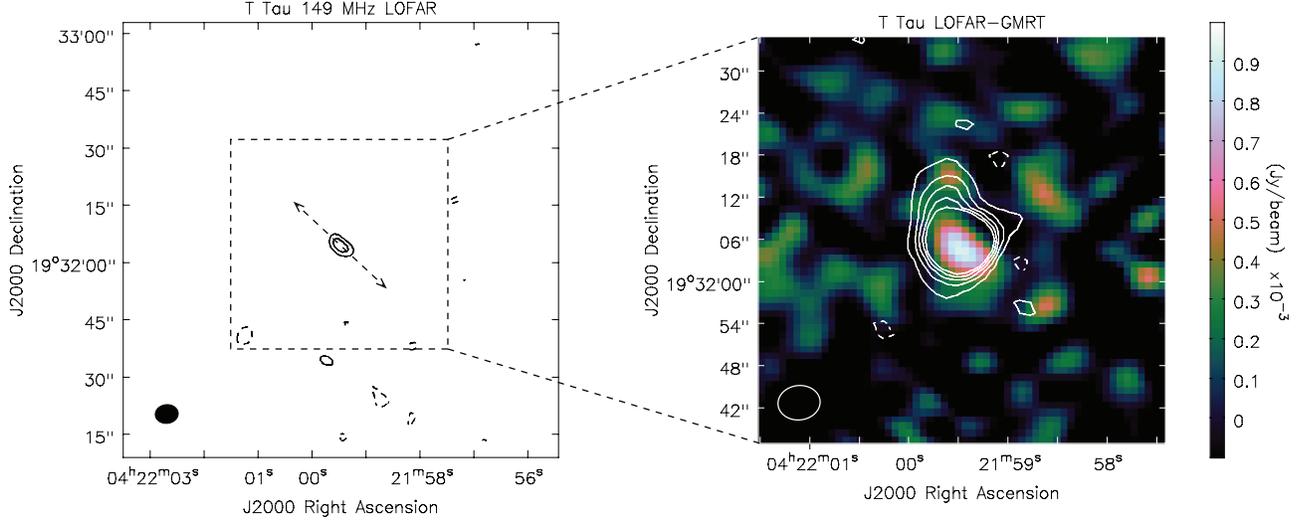}
\caption{Left: The LOFAR image of the T~Tau system at 149\,MHz with a \textsc{CLEAN} beam of $6\farcs01 \times 4\farcs90$ and position angle $276\fdg16$ shown by the filled ellipse. Contours are shown at -3 (dashed), 3 and $4\,\sigma_{\rm rms}$, where $\sigma_{\rm rms}$ corresponds to 0.20\,mJy\,beam$^{-1}$. The dashed line with arrows indicates the axis of the T~Tau~Sb extended component seen at higher frequencies \citep[with position angle of $47^\circ$, see][]{Loinard2007a} with which the LOFAR detection is consistent. The emission has a peak flux of 0.95\,mJy\,beam$^{-1}$ and an integrated flux of $1.75 \pm 0.36$\,mJy. Right: The colour scale shows the 149\,MHz LOFAR data. Contours are the GMRT data of \citet{Ainsworth2016} at 608\,MHz, shown at -3 (dashed), 3, 6, 9, 12 and $15\,\sigma_{\rm rms}$ where $\sigma_{\rm rms}$ corresponds to 45\,$\mu$Jy\,beam$^{-1}$. The beam size of the 608\,MHz GMRT image is $6\farcs03 \times 5\farcs01$ with position angle $276\fdg18$ which is only 3\% greater than the LOFAR beam area.}
\label{fig:ttau}
\end{center}
\end{figure*}

Figure~\ref{fig:ttau} (left) shows the LOFAR detection of the T~Tau system with a peak flux of $S_{\rm 149\,GHz,peak} = 0.96 \pm 0.20$\,mJy\,beam$^{-1}$ at a significance of $4.8\,\sigma_{\rm rms}$ excluding systematic errors. The integrated flux is $S_{\rm 149\,GHz,int} = 1.75 \pm 0.36$\,mJy (including the systematic contribution discussed in Section~\ref{sec:observations}) with a Gaussian fit of $(9\farcs29 \pm 2\farcs32)\times (5\farcs80 \pm 1\farcs04)$ and a position angle of $46\fdg8\pm23\fdg7$. When compared to the restoring beam of $6\farcs01 \times 4\farcs90$, this indicates that the emission has been partially resolved at least along one direction. The deconvolved source size was $7\farcs60 \times 1\farcs41$ with a position angle of $40\fdg5$. Figure~\ref{fig:ttau} (right) shows the LOFAR data in colour with the contours overlaid from the GMRT observations of \citet{Ainsworth2016} at 608\,MHz. The radio emission appears extended along the same direction at both frequencies and consistent with the position angle of $47^\circ$ of the extended component observed by \citet{Loinard2007a} at 8\,GHz. As the T~Tau system is located at a distance of 148\,pc \citep{Loinard2007b}, LOFAR's resolution of $\simeq 5\farcs4$ means that the T~Tau system is being probed on a scale of several hundred au. At this resolution, we expect the radio emission to be heavily dominated by T~Tau~S (specifically the extended, free--free component from T~Tau~Sb) based on the results of \citet{Johnston2003} and \citet[][see Section~\ref{sec:introduction} and Figure~\ref{fig:ttau_var}]{Loinard2007a}, with which indeed the LOFAR detection appears to be consistent.

\section{Discussion}
\label{sec:discussion}

\subsection{SED Analysis}
\label{sec:sed_analysis}

\begin{figure*}
\begin{center}

\subfloat{
\includegraphics[width=0.5\textwidth]{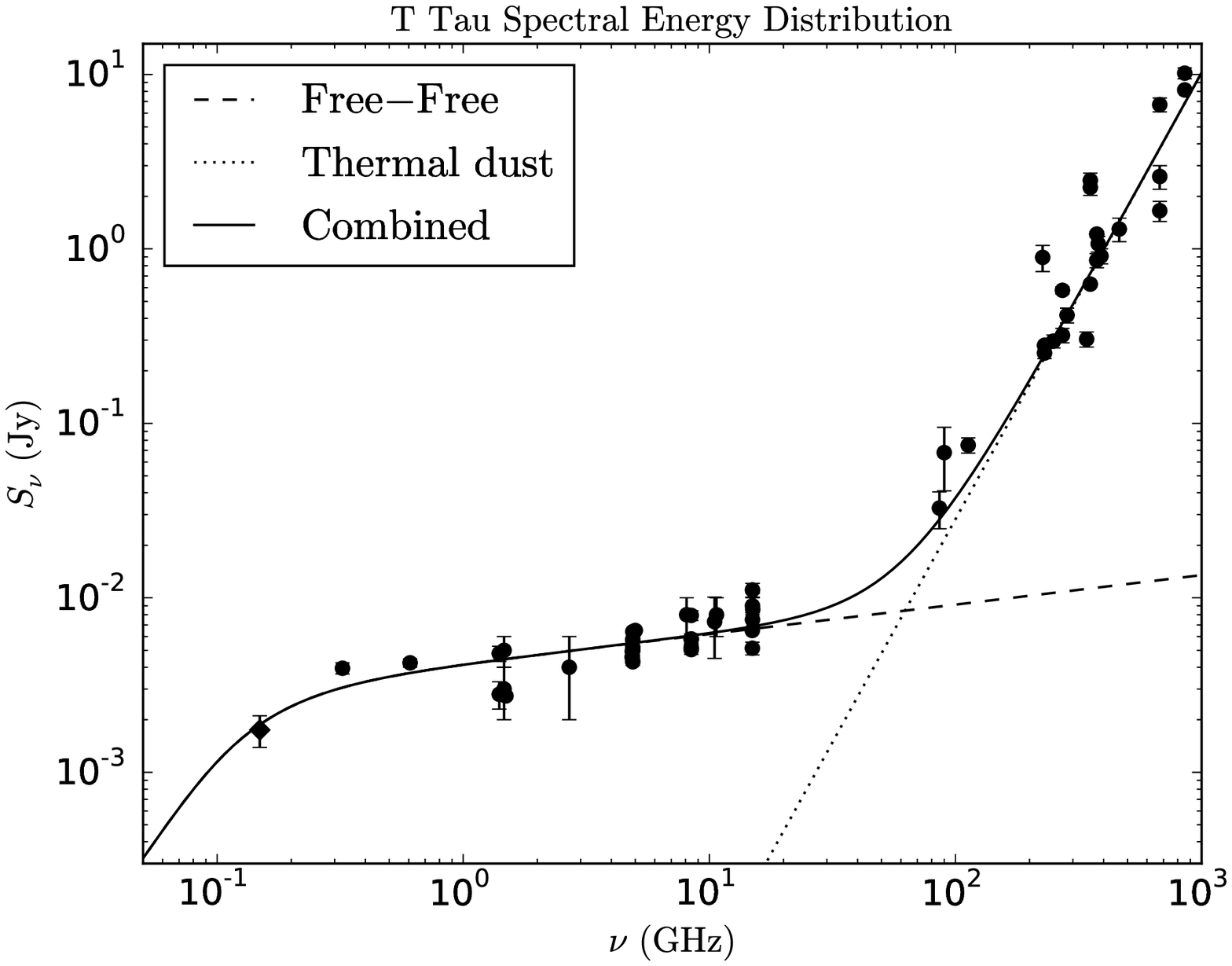}
}
\subfloat{
\includegraphics[width=0.5\textwidth]{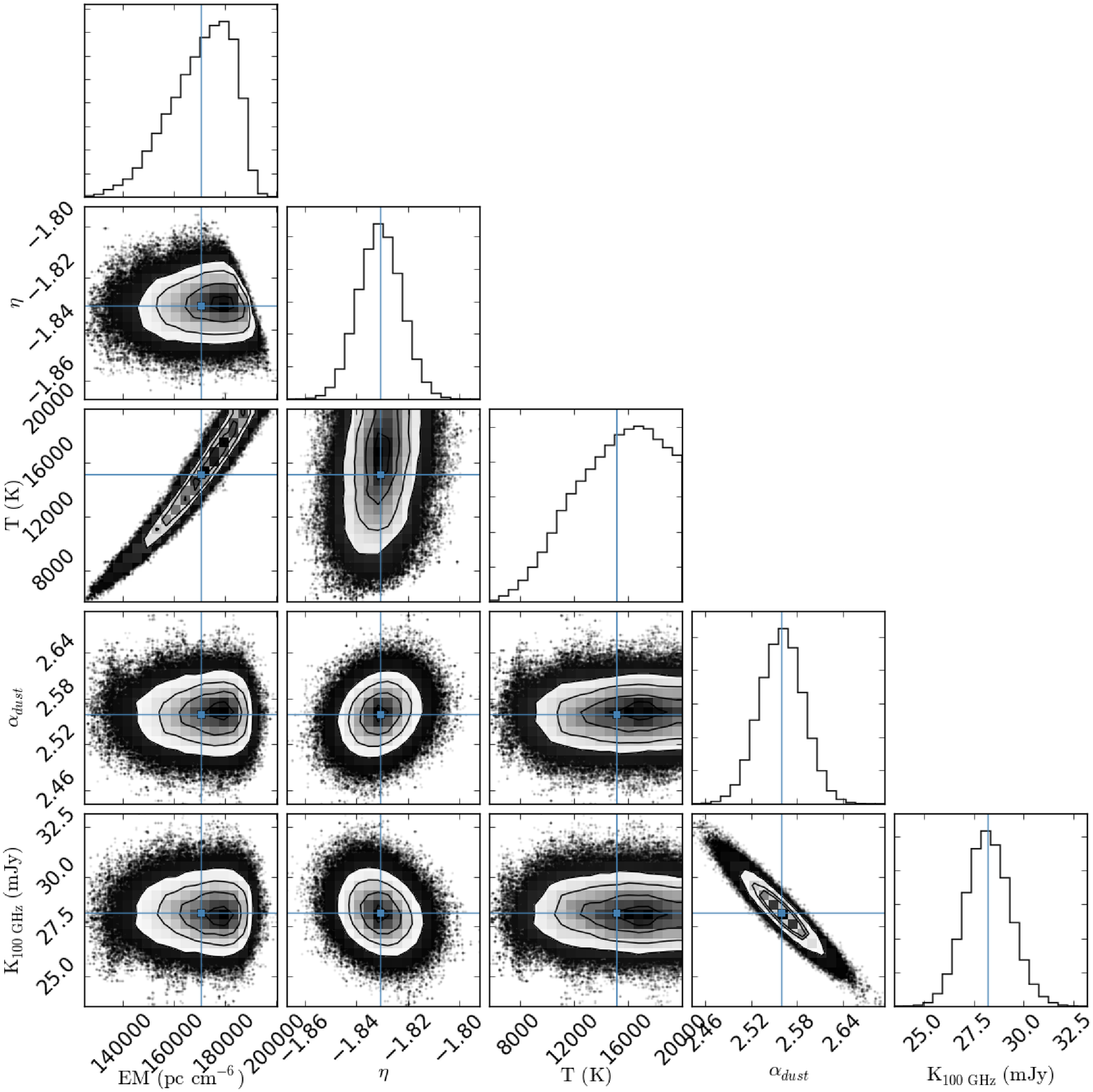}
}

\caption{The spectral energy distribution for T~Tau (left) and plots for fitted parameters (right). The measured flux density from this work is shown as a filled diamond and those from the literature are shown as filled circles \citep[see][]{Ainsworth2016}. The SED is fitted with a combined model to describe the free--free emission from the outflow (dashed line) and the thermal dust emission from the circumstellar disk/envelope (dotted line). The total fit is shown as a continuous line. The corner plot on the right shows the posterior distributions calculated during the MCMC fitting for the emission measure, temperature and spectral index of the free--free emission, as well as the normalisation constant and spectral index of the power law describing the thermal dust emission. The histograms represent one dimensional distributions of the variable in question, while the contour plots show two dimensional distributions of two variables with contours at 0.5, 1.0, 1.5 and $2.0\,\sigma$. The lines indicate the position of the mean values (see Section~\ref{sec:discussion}).}
\label{fig:spectrum}
\end{center}
\end{figure*}

The SED for the T~Tau system is presented in Figure~\ref{fig:spectrum}, which combines the flux density measured with LOFAR in this work and the data used in \citet{Ainsworth2016}. These authors ensured that integrated fluxes in the free--free, low-frequency part of the spectrum ($\nu \lesssim 15$\,GHz) were taken from observations with a similar angular resolution so that only emission on similar spatial scales is modelled. LOFAR and the GMRT do not resolve T~Tau~N and S, therefore where higher resolution VLA data is used \citep[e.g.][]{Skinner1994, Johnston2003, Loinard2007a}, the T~Tau~N and S components have been added together and as a result may still have some missing flux. This was also done, in so far as was possible, for the high frequency  ($\nu \gtrsim 15$\,GHz) data. 

There are two distinct regimes in the SED which need to be modelled: free--free emission associated with the partially ionised outflows at low frequencies and thermal dust emission associated with the circumstellar disks at high frequencies \citep[see discussion in][]{Ainsworth2016}. Although we do not expect the higher frequency data to contaminate the flux densities at very low frequencies, the long wavelength tail of the thermal dust emission has been shown to contribute to the centimetre-wave flux densities of YSOs which can affect the spectral modelling of the entire free--free spectrum and must therefore be included \citep{Scaife2012b}.

As the turnover in the free--free spectrum had not yet been detected at GMRT frequencies, \citet{Ainsworth2016} modelled the partially optically thin/thick free--free emission using a single power law of $\alpha=0.17\pm0.01$. However, the flux density measured at 149\,MHz with LOFAR lies significantly ($3\,\sigma$) below this power law (see Figure~\ref{fig:spectrum}). Assuming this drop in flux is due to the transition between optically thin and thick behaviour in the free--free spectrum, we now have the spectral constraint necessary to begin to model the full free--free spectrum. The radio flux density from such emission is given by
\begin{eqnarray}
\left( \frac{S_{\nu}}{\rm mJy} \right) & = & 7.21586\times 10^{-4} \left(\frac{\nu}{\rm GHz} \right)^{2} \times \nonumber \\
 & & \left( \frac{T_{\rm e}}{\rm K}\right) \left( 1 - e^{-\tau_{\nu}} \right) \left(\frac{\Omega_{s}}{\rm arcsec^{2}}\right) \label{eq:ff}
\end{eqnarray}
\citep[see e.g.][]{Scaife2013}, where $\tau_\nu$ is the optical path length (depth) for free--free emission which can be approximated by 
\begin{equation}
\tau_{\nu} = 8.235 \times 10^{-2} \left( \frac{T_{e}}{\rm K} \right)^{-1.35} \left( \frac{\nu}{\rm GHz} \right)^{-2.1} \left( \frac{EM}{\rm pc\,cm^{-6}} \right) \label{eq:tau}
\end{equation}
\citep{Altenhoff1960}. This approximation has an estimated accuracy of over 94\% for LOFAR frequencies and temperatures of up to $1.5 \times 10^{4}$\,K. In these equations, $T_{\rm e}$ is the electron temperature and $\Omega_{\rm s} = \frac{\pi \theta^{2}}{4 \ln 2}$ corresponds to the solid angle of the emission where $\theta$ is the geometric mean of the deconvolved angular size. $EM$ is the characteristic property of emission measure which is a function of the average electron number density ($n_{\rm e}$) and source size \citep{Mezger1967},  
\begin{equation}
\left(\frac{EM}{\rm pc\,cm^{-6}}\right)  =  7.1\times10^{-3} \left(\frac{D}{{\rm kpc}}\right) \left(\frac{\theta}{\rm arcsec}\right) \left(\frac{n_{\rm e}}{\rm cm^{-3}}\right)^2 \label{eq:em}
\end{equation}
where $D$ is the distance to the source.

We model the flux distribution over the entire range of frequencies using a combination of equation~\ref{eq:ff} to model the free--free emission at low frequencies and a power law to model the thermal dust emission at high frequencies. This model has the form
\begin{eqnarray}
\left( \frac{S_{\nu}}{\rm mJy} \right) & = & 7.21586\times 10^{-4} \times \nonumber \\
 & &  \left(\frac{\nu}{\rm GHz} \right)^{2} \left( \frac{T_{\rm e}}{\rm K}\right) \left( 1 - e^{-\tau_{\nu}} \right) \left(\frac{\Omega_{s}}{\rm arcsec^{2}}\right) \nonumber   \\
 & & + K_{\rm 100\,GHz} \left(\frac{\nu}{\rm 100\,GHz} \right)^{\alpha_{\rm dust}} \label{eq:model}
\end{eqnarray}
where $\alpha_{\rm dust}$ is the power law index to be fitted to the high frequency data and $K_{\rm 100\,GHz}$ is a normalisation constant which normalises the high frequency component of the model at 100\,GHz.

In a fully optically thick medium ($\tau_{\nu} \gg 1$) the observed flux density scales with frequency as $S_{\nu} \propto \nu^{2} $, while in an optically thin medium ($1-e^{\tau_{\nu}} \approx \tau_{\nu}$) $S_{\nu} \propto \nu^{-0.1}$ \citep[see e.g.][]{Scaife2013}. Although in the case of the T~Tau system, where a combination of emission from three distinct outflow sources is detected in a medium that may be partially optically thick, or have contributions to the total flux from both thermal and non-thermal emission, it might be expected that the overall dependence of flux density on frequency is more complicated. 

As discussed in Section~\ref{sec:introduction}, T~Tau~S has been shown to dominate the total radio flux density and spectral index over the entire system. However, the steeper ($\alpha\simeq1$) thermal emission from T~Tau~N will add a small contribution to the total spectral index of the unresolved T~Tau~N+S system. Using a least-squares fitting method in Python, the spectral index of the combined T~Tau~N+S system is expected to be of order $\alpha\simeq0.12$ based on the VLA observations of \citet[][see Appendix~\ref{app:si}]{Johnston2003}. We therefore replace the $\tau_\nu\propto\nu^{-2.1}$ dependence (equation~\ref{eq:tau}) with $\tau_\nu\propto\nu^{\eta}$ in our model to account for this small contribution from T~Tau~N. This will also allow for intermediate values of the optical depth from regions where the electron density distribution is different to $n_{\rm e}(r)\propto r^{-2}$ \citep[i.e. exhibits a more jet-like geometry, see e.g.][]{Reynolds1986} and/or has contributions from non-thermal, gyrosynchrotron emission due to magnetic activity. We note, however, that the gyrosynchrotron radiation has been shown to only dominate the emission detected on small ($\ll1$\,au) scales \citep{Smith2003, Loinard2007a, Loinard2007b} and we expect the total radio emission on large ($\sim100$\,au) scales to be predominantly associated with free--free radiation due to shock ionisation from the outflow driven by T~Tau~Sb \citep{Loinard2007a}.

The Python Monte Carlo analysis package, \textsc{pymc} \citep{Patil2010}, was used to find the best fit between equation~\ref{eq:model} and the data in Figure~\ref{fig:spectrum} using a Markov Chain Monte Carlo (MCMC) approach. The prior ranges used for fitting were $-2.1 \leqslant \eta \leqslant 0$, $ 3\times10^{3}\leqslant \frac{T_{e}}{\rm K} \leqslant 2\times 10^{4}$, $ 10^{4} \leqslant \frac{EM}{\rm pc\,cm^{-6}} \leqslant 10^{7}$, $10 \leqslant \frac{K_{\rm 100\,GHz}}{\rm mJy} \leqslant 40$ and $0 \leqslant \alpha_{\rm dust} \leqslant 4$. The fitting also allowed the posterior distributions of the fitted parameters to be determined, as well the uncertainties in their mean values (see Figure~\ref{fig:spectrum}).

\subsection{Constraints on physical parameters}
\label{sec:constraints}

Through a measurement of the optically thick surface due to the detection of the turnover frequency, we can break the degeneracy between source size and electron density inherent within the emission measure and constrain the mass of the ionised gas surrounding the T~Tau system on scales of several hundreds of au. Figure~\ref{fig:spectrum} shows the fitted value of $EM=(1.67 \pm 0.14 )\times10^5$\,pc\,cm$^{-6}$ to be well constrained by our SED modelling and we measure a geometric mean of the deconvolved angular source size of $\theta=3\farcs27$ from our LOFAR image. We then use equation~\ref{eq:em} to find an average electron number density of $n_{\rm e}=(7.2 \pm 2.1)\times10^{3}$\,cm$^{-3}$ and an approximate mass may be derived from this result. Assuming a simple spherical geometry, the density distribution can be integrated to give the ionised gas mass as $M_{\rm ion}=\frac{4}{3}\pi r_{\rm gas}^{3}m_{\rm H}n_{\rm e}=(1.0 \pm 1.8)\times10^{-6}$\,M$_{\odot}$, where the radius of the sphere $r_{\rm gas} = D \frac{\theta}{2}$ and $m_{\rm H}$ is the atomic mass of Hydrogen.

\begin{figure}
\begin{center}
\includegraphics[width=\columnwidth]{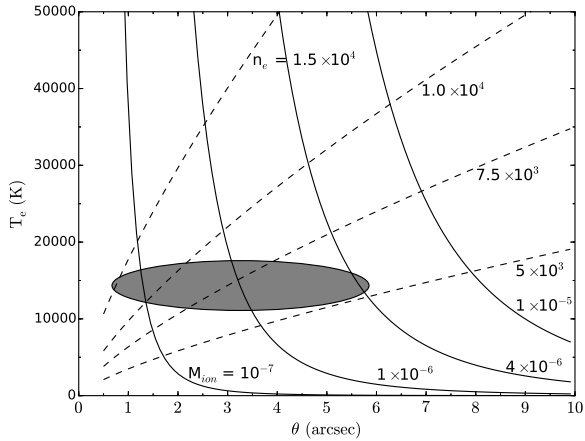}
\caption{Selected values of electron density ($n_{e}$ in \,cm$^{-3}$) and total ionised mass ($M_{\rm ion}$ in solar masses) plotted against electron temperature ($T_{\rm e}$) and source size ($\theta$) using equations \ref{eq:tau} and \ref{eq:em} and assuming a spherical cloud of constant density. The shaded region indicates the area within 1\,$\sigma$ of the geometric mean of the deconvolved angular source size and fitted temperature ($\theta = 3\farcs27 \pm 2\farcs59$, $T_{\rm e} = (1.4 \pm 0.3) \times 10^{4}$\,K).}
\label{fig:gas-density-mass}
\end{center}
\end{figure}

The model fitted returns an electron temperature of $T_{\rm e} = (1.4 \pm 0.3) \times 10^{4}$\,K for the low frequency component, a value consistent with the temperature of $10^{4}$\,K used in the modelling of \citet{Ainsworth2016}. It is clear from Figure~\ref{fig:spectrum} that this estimate does not constrain the temperature very well as a large range of temperatures from about $10^{4}$ to $1.8 \times 10^{4}$\,K fit the data almost equally well, thus the fitted value can be only regarded as a rough estimate. Figure \ref{fig:gas-density-mass} shows how the electron density and total ionised mass vary with deconvolved source size and temperature according to equations \ref{eq:tau} and \ref{eq:em} for a spherical gas cloud of constant density. The shaded region indicates the $1\,\sigma$ boundary around the measured deconvolved source size and fitted electron temperature reported above.

The high frequency component was fitted with values of $K_{\rm 100\,GHz} = 28.3 \pm  1.1$\,mJy and $\alpha_{\rm dust} = 2.56 \pm 0.03$ (corresponding to a dust opacity index $\beta \approx \alpha_{\rm dust} - 2 = 0.56$). The posterior distributions plotted in Figure~\ref{fig:spectrum} show these results to be very well constrained and, as expected, closely matched with the original high frequency results of \citet{Ainsworth2016}. 

A turnover frequency of $157 \pm 27$\,MHz was calculated from equation~\ref{eq:tau} using the model parameters and setting $\tau_\nu = 1$. This value is consistent with the interpretation that the low frequency turnover of the free--free spectrum has been observed with LOFAR. A mean value of $-1.83\pm0.01$ was determined for $\eta$, the modified spectral index term in equation~\ref{eq:tau}. Thus the flux density scales with $\nu^{0.17}$ above the turnover frequency of 157\,MHz. This is consistent with the spectral index found in \citet{Ainsworth2016} and does not differ greatly from the canonical relationship of $\nu^{-0.1}$ for optically thin emission, but takes into account the partially optically thick nature of the plasma and the contribution from T~Tau~N (see Appendix~\ref{app:si}).  

The spectral index will also be affected by the variability of T~Tau~S, which can be clearly seen in Figures~\ref{fig:ttau_var} and \ref{fig:spectrum} by the range of flux densities for a given frequency measured at different epochs. It is possible that the significant decrease in flux density at 149\,MHz is due to variability, as the flux density has been shown to change by up to 50\% in the past \citep[see e.g.][]{Johnston2003, Loinard2007a} and an increase in the flux by up to 50\% would be consistent with the $S_{\nu}\propto\nu^{0.17}$ spectrum. 

The variability has been suggested to be due to the compact, non-thermal component associated with T~Tau~Sb and a range of emission mechanisms have been proposed to produce it, the most likely of which being non-thermal gyrosynchrotron emission \citep{Skinner1994, Johnston2003, Smith2003, Loinard2007a}. The spectral index for this component was found to be $\alpha\simeq0.5$ by \citet{Loinard2007a}, which suggests that the peak of the gyrosynchroton spectrum is at higher frequencies and therefore we might expect to see less variability at lower frequencies. This is supported by the multi-epoch, multi-frequency observations of \citet{Johnston2003} which show that the T~Tau~S flux density increases approximately by a factor of four between 1986 and 1988 at 15\,GHz and only by a factor of approximately two at 5\,GHz (see Figure~\ref{fig:ttau_var}). However, the spectral index of this component has been shown to be variable and helicity reversals of the circular polarisation have been observed \citep[see e.g.][]{Skinner1994, Smith2003}. It is therefore difficult to quantify the relative contribution of this component to the total flux density and variability at LOFAR frequencies.

In the absence of multi-epoch data at these very low frequencies, the LOFAR data suggests a turnover in the free--free spectrum. Simultaneous multi-frequency data extending from LOFAR to VLA frequencies can be used to probe the variability in this regime and constrain these physical properties further.

\section{Conclusions}
\label{sec:conclusions}

The T~Tau system has been successfully detected at 149\,MHz with LOFAR. We suggest that this emission is dominated by the extended, free--free component proposed to arise from shock ionisation in the outflow of T~Tau~Sb. This is the lowest frequency observation of a YSO to date and the first detection of a YSO with LOFAR. The integrated flux of $1.75 \pm 0.36$\,mJy lies significantly below the power law extrapolated from \citet{Ainsworth2016}. Additional LOFAR observations are required to probe the variability of the flux density at 149\,MHz. In the absence of such multi-epoch observations, we suggest that the turnover in the free--free spectrum has been observed. This flux measurement, along with an estimate of the size of the emitting region based on the partially resolved detection, has allowed the degeneracy between emission measure and electron density to be broken by fitting the SED between 149\,MHz and 900\,GHz. New estimates of the emission measure, electron density and ionised gas mass have been made. This result establishes the utility of LOFAR as a probe of the spectra of YSOs close to their free--free turnover points, however sensitivity constraints mean that, for now, only the brightest YSOs will make suitable targets for observation at these low frequencies.

\appendix

\section{Predicted Spectral Behaviour of T~Tau~N+S System}
\label{app:si}

To ensure that only emission on similar spatial scales is modelled, the SED presented in Figure~\ref{fig:spectrum} comprises integrated fluxes taken from observations with similar angular resolutions to these LOFAR observations. However, the LOFAR data presented cannot resolve the T~Tau~N and S sources and therefore the SED will have contributions from each. 

To estimate the overall spectral index above the turnover frequency for the unresolved T~Tau~N+S system, we fit the sum of two power laws based on the time averaged data presented in \citet[][see Figure~\ref{fig:ttau_var}]{Johnston2003} using least-squares minimisation in Python. The power laws are of the form
\begin{equation}
S_\nu = A_{\rm J,avg} \left( \frac{\nu}{\rm 8\,GHz} \right)^{\alpha_{\rm J,avg}}.
\end{equation}
$A_{\rm J, avg}$ is the 8\,GHz flux density averaged over the five epochs measured in \citet{Johnston2003} and equals 0.76\,mJy for T~Tau~N and 5.16\,mJy for T~Tau~S. $\alpha_{\rm J,avg}$ is the spectral index calculated between 5 and 15\,GHz (except for epochs 1989 and 2001 which were calculated between 8 and 15\,GHz) and averaged over the seven epochs measured in \citet{Johnston2003}. The average spectral index values are 1.04 for T~Tau~N and 0.03 for T~Tau~S. We note that the T~Tau~S spectrum itself will have contributions from both the compact/non-thermal and extended/thermal components associated with gyrosynchrotron radiation from the T~Tau~Sb magnetic fields and free--free radiation from shock ionised plasma of the T~Tau~Sb outflow, respectively. However, these components were unresolved by \citet{Johnston2003} and therefore we do not model them separately here.

\begin{figure}
\begin{center}
\includegraphics[width=0.5\textwidth]{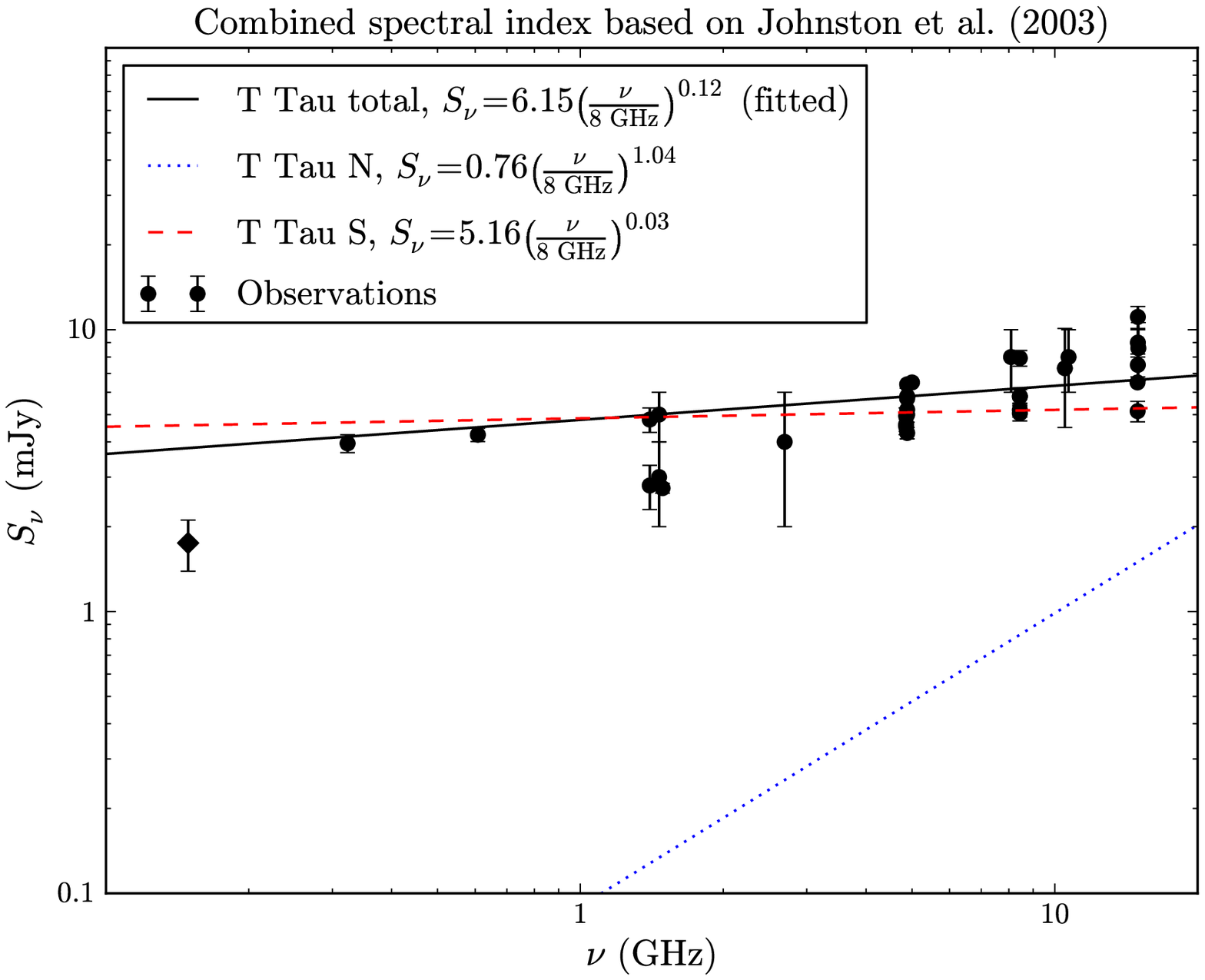}
\caption{The predicted power law spectrum for the unresolved T~Tau~N+S system based on the data presented in \citet{Johnston2003}. The (blue) dotted line is the T~Tau~N spectrum where 0.76\,mJy is the time averaged flux density at 8\,GHz and 1.04 is the time averaged spectral index. The (red) dashed line is the T~Tau~S spectrum where 5.16\,mJy is the time averaged flux density at 8\,GHz and 0.03 is the time averaged spectral index. The (black) solid line is the fit to the sum of the two power laws using least-squares minimisation in Python with normalised flux density of 6.15\,mJy at 8\,GHz and an spectral index of 0.12, representative of the T~Tau~N+S system. Archival observations are shown as filled circles and the LOFAR measurement is a filled diamond, which is clearly discrepant from the fitted spectrum.}
\label{fig:Johnston}
\end{center}
\end{figure}

The two spectra were then added together and fitted with a single power law, resulting in an overall spectrum with $S_{\nu}\propto\nu^{0.12}$, see Figure~\ref{fig:Johnston}. This frequency dependence is close to the $\nu^{0.17}$ measured in Section~\ref{sec:constraints} and fits the archival flux density measurements (filled circles in Figure~\ref{fig:Johnston}). The LOFAR datum (filled diamond) is clearly discrepant from this power law and, if not due to variability, suggests we have detected the turnover in the free--free spectrum.

\section*{Acknowledgements}

LOFAR, the Low Frequency Array designed and constructed by ASTRON, has facilities in several countries, that are owned by various parties (each with their own funding sources), and that are collectively operated by the International LOFAR Telescope (ILT) foundation under a joint scientific policy. The authors wish to acknowledge the DJEI/DES/SFI/HEA Irish Centre for High-End Computing (ICHEC) for the provision of computational facilities and support.  CPC, REA and TPR would like to acknowledge support from Science Foundation Ireland under grant 13/ERC/I2907. AMS gratefully acknowledges support from the European Research Council under grant  ERC-2012-StG-307215 LODESTONE. SC acknowledges financial support from the UnivEarthS Labex program of Sorbonne Paris Cit\'e (ANR-10-LABX-0023 and ANR-11-IDEX-0005-02).

\software{AOFlagger \citep{Offringa2012}, AWImager \citep{Tasse2013}, CASA \citep{McMullin2007}, Prefactor \citep{Weeren2016}, PyBDSM \citep{Mohan2015}, PyMC \citep{Patil2010}, SAGECal \citep{Kazemi2011,Kazemi2013}}

\bibliographystyle{apj}
\bibliography{T_Tau_LOFAR_abbreviated.bib}

\begin{thebibliography}{}
\expandafter\ifx\csname natexlab\endcsname\relax\def\natexlab#1{#1}\fi

\bibitem[{Ainsworth {et~al.}(2016)Ainsworth, Scaife, Green, Coughlan, \&
  Ray}]{Ainsworth2016}
Ainsworth, R.~E., Scaife, A. M.~M., Green, D.~A., Coughlan, C.~P., \& Ray,
  T.~P. 2016, MNRAS, 459, 1248

\bibitem[{Ainsworth {et~al.}(2014)Ainsworth, Scaife, Ray, Taylor, Green, \&
  Buckle}]{Ainsworth2014}
Ainsworth, R.~E., Scaife, A. M.~M., Ray, T.~P., {et~al.} 2014, ApJ, 792, L18

\bibitem[{Altenhoff {et~al.}(1960)Altenhoff, Mezger, Strassl, Wendker, \&
  Westerhout}]{Altenhoff1960}
Altenhoff, W., Mezger, P.~G., Strassl, H., Wendker, H., \& Westerhout, G. 1960,
  Veroff Sternwarte, Bonn, 48

\bibitem[{Anglada {et~al.}(2015)Anglada, Rodr{\'{i}}guez, \&
  Carrasco-Gonzalez}]{Anglada2015}
Anglada, G., Rodr{\'{i}}guez, L.~F., \& Carrasco-Gonzalez, C. 2015, Proceedings
  of Advancing Astrophysics with the Square Kilometre Array (AASKA14). 9 -13
  June

\bibitem[{Carrasco-Gonz{\'{a}}lez {et~al.}(2010)Carrasco-Gonz{\'{a}}lez,
  Rodr{\'{i}}guez, Anglada, Mart{\'{i}}, Torrelles, \&
  Osorio}]{Carrasco-Gonzalez2010}
Carrasco-Gonz{\'{a}}lez, C., Rodr{\'{i}}guez, L.~F., Anglada, G., {et~al.}
  2010, Sci, 330, 1209

\bibitem[{Cohen {et~al.}(2007)Cohen, Lane, Cotton, Kassim, Lazio, Perley,
  Condon, \& Erickson}]{Cohen2007}
Cohen, A.~S., Lane, W.~M., Cotton, W.~D., {et~al.} 2007, AJ, 134, 1245

\bibitem[{Condon {et~al.}(1998)Condon, Cotton, Greisen, Yin, Perley, Taylor, \&
  Broderick}]{Condon1998}
Condon, J.~J., Cotton, W.~D., Greisen, E.~W., {et~al.} 1998, AJ, 115, 1693

\bibitem[{Dyck {et~al.}(1982)Dyck, Simon, \& Zuckerman}]{Dyck1982}
Dyck, H.~M., Simon, T., \& Zuckerman, B. 1982, ApJ, 255, L103

\bibitem[{Dzib {et~al.}(2015)Dzib, Loinard, Rodr{\'{i}}guez, Mioduszewski,
  Ortiz-Le{\'{o}}n, Kounkel, Pech, Rivera, Torres, Boden, Hartmann, Evans,
  Brice{\~{n}}o, \& Tobin}]{Dzib2014}
Dzib, S.~A., Loinard, L., Rodr{\'{i}}guez, L.~F., {et~al.} 2015, ApJ, 801,
  arXiv:1412.6445

\bibitem[{Gustafsson {et~al.}(2010)Gustafsson, Kristensen, Kasper, \&
  Herbst}]{Gustafsson2010}
Gustafsson, M., Kristensen, L.~E., Kasper, M., \& Herbst, T.~M. 2010, A{\&}A,
  517, A19

\bibitem[{Intema {et~al.}(2016)Intema, Jagannathan, Mooley, \&
  Frail}]{Intema2016}
Intema, H.~T., Jagannathan, P., Mooley, K.~P., \& Frail, D.~A. 2016, Submitted
  to A{\&}A. eprint arXiv:1603.04368

\bibitem[{{Johnston} {et~al.}(2004){Johnston}, {Fey}, {Gaume}, {Hummel},
  {Garrington}, {Muxlow}, \& {Thomasson}}]{Johnston2004}
{Johnston}, K.~J., {Fey}, A.~L., {Gaume}, R.~A., {et~al.} 2004, \apjl, 604, L65

\bibitem[{Johnston {et~al.}(2003)Johnston, Gaume, Fey, de~Vegt, \&
  Claussen}]{Johnston2003}
Johnston, K.~J., Gaume, R.~A., Fey, A.~L., de~Vegt, C., \& Claussen, M.~J.
  2003, AJ, 125, 858

\bibitem[{Joy(1945)}]{Joy1945}
Joy, A.~H. 1945, ApJ, 102, 168

\bibitem[{Kazemi \& Yatawatta(2013)}]{Kazemi2013}
Kazemi, S., \& Yatawatta, S. 2013, MNRAS, 435, 597

\bibitem[{Kazemi {et~al.}(2011)Kazemi, Yatawatta, Zaroubi, Lampropoulos,
  de~Bruyn, Koopmans, \& Noordam}]{Kazemi2011}
Kazemi, S., Yatawatta, S., Zaroubi, S., {et~al.} 2011, MNRAS, 414, 1656

\bibitem[{{K{\"o}hler} {et~al.}(2008){K{\"o}hler}, {Ratzka}, {Herbst}, \&
  {Kasper}}]{Kohler2008}
{K{\"o}hler}, R., {Ratzka}, T., {Herbst}, T.~M., \& {Kasper}, M. 2008, \aap,
  482, 929

\bibitem[{Koresko(2000)}]{Koresko2000}
Koresko, C.~D. 2000, ApJ, 531, L147

\bibitem[{Lane {et~al.}(2012)Lane, Cotton, Helmboldt, \& Kassim}]{Lane2012}
Lane, W.~M., Cotton, W.~D., Helmboldt, J.~F., \& Kassim, N.~E. 2012, RaSc, 47,
  n/a

\bibitem[{{Loinard} {et~al.}(2005){Loinard}, {Mioduszewski},
  {Rodr{\'{\i}}guez}, {Gonz{\'a}lez}, {Rodr{\'{\i}}guez}, \&
  {Torres}}]{Loinard2005}
{Loinard}, L., {Mioduszewski}, A.~J., {Rodr{\'{\i}}guez}, L.~F., {et~al.} 2005,
  \apjl, 619, L179

\bibitem[{{Loinard} {et~al.}(2007{\natexlab{a}}){Loinard}, {Rodr{\'{\i}}guez},
  {D'Alessio}, {Rodr{\'{\i}}guez}, \& {Gonz{\'a}lez}}]{Loinard2007a}
{Loinard}, L., {Rodr{\'{\i}}guez}, L.~F., {D'Alessio}, P., {Rodr{\'{\i}}guez},
  M.~I., \& {Gonz{\'a}lez}, R.~F. 2007{\natexlab{a}}, \apj, 657, 916

\bibitem[{{Loinard} {et~al.}(2007{\natexlab{b}}){Loinard}, {Torres},
  {Mioduszewski}, {Rodr{\'{\i}}guez}, {Gonz{\'a}lez-L{\'o}pezlira}, {Lachaume},
  {V{\'a}zquez}, \& {Gonz{\'a}lez}}]{Loinard2007b}
{Loinard}, L., {Torres}, R.~M., {Mioduszewski}, A.~J., {et~al.}
  2007{\natexlab{b}}, \apj, 671, 546

\bibitem[{McMullin {et~al.}(2007)McMullin, Waters, Schiebel, Young, \&
  Golap}]{McMullin2007}
McMullin, J.~P., Waters, B., Schiebel, D., Young, W., \& Golap, K. 2007,
  Astronomical Data Analysis Software and Systems XVI ASP Conference Series,
  376

\bibitem[{Mezger \& Henderson(1967)}]{Mezger1967}
Mezger, P.~G., \& Henderson, A.~P. 1967, ApJ, 147, 471

\bibitem[{Mohan \& Rafferty(2015)}]{Mohan2015}
Mohan, N., \& Rafferty, D. 2015, ASCL

\bibitem[{Offringa {et~al.}(2012)Offringa, van~de Gronde, \&
  Roerdink}]{Offringa2012}
Offringa, A.~R., van~de Gronde, J.~J., \& Roerdink, J. B. T.~M. 2012, A{\&}A,
  539, A95

\bibitem[{{Padovani} {et~al.}(2016){Padovani}, {Marcowith}, {Hennebelle}, \&
  {Ferri{\`e}re}}]{Padovani2016}
{Padovani}, M., {Marcowith}, A., {Hennebelle}, P., \& {Ferri{\`e}re}, K. 2016,
  \aap, 590, A8

\bibitem[{Patil {et~al.}(2010)Patil, Huard, \& Fonnesbeck}]{Patil2010}
Patil, A., Huard, D., \& Fonnesbeck, C.~J. 2010, Journal of statistical
  software, 35, 1

\bibitem[{Ray {et~al.}(1997)Ray, Muxlow, Axon, Brown, Corcoran, Dyson, \&
  Mundt}]{Ray1997}
Ray, T.~P., Muxlow, T. W.~B., Axon, D.~J., {et~al.} 1997, Natur, 385, 415

\bibitem[{Reipurth {et~al.}(1997)Reipurth, Bally, \& Devine}]{Reipurth1997}
Reipurth, B., Bally, J., \& Devine, D. 1997, AJ, 114, 2708

\bibitem[{Rengelink {et~al.}(1997)Rengelink, Tang, de~Bruyn, Miley, Bremer,
  R�ttgering, \& Bremer}]{Rengelink1997}
Rengelink, R.~B., Tang, Y., de~Bruyn, A.~G., {et~al.} 1997, A {\&} AS, 124, 259

\bibitem[{{Reynolds}(1986)}]{Reynolds1986}
{Reynolds}, S.~P. 1986, \apj, 304, 713

\bibitem[{{Scaife}(2011)}]{Scaife2011}
{Scaife}, A.~M.~M. 2011, The Astronomer's Telegram, 3786

\bibitem[{Scaife(2013)}]{Scaife2013}
Scaife, A. M.~M. 2013, AdAst, 2013, 1

\bibitem[{Scaife \& Heald(2012)}]{Scaife2012a}
Scaife, A. M.~M., \& Heald, G.~H. 2012, MNRAS: Letters, 423, L30

\bibitem[{Scaife {et~al.}(2012)Scaife, Buckle, Ainsworth, Davies, Franzen,
  Grainge, Hobson, Hurley-Walker, Lasenby, Olamaie, Perrott, Pooley, Ray,
  Richer, Rodr{\'{i}}guez-Gonz{\'{a}}lvez, Saunders, Schammel, Scott, Shimwell,
  Titterington, \& Waldram}]{Scaife2012b}
Scaife, A. M.~M., Buckle, J.~V., Ainsworth, R.~E., {et~al.} 2012, MNRAS, 420,
  3334

\bibitem[{{Schwartz} {et~al.}(1986){Schwartz}, {Simon}, \&
  {Campbell}}]{Schwartz1986}
{Schwartz}, P.~R., {Simon}, T., \& {Campbell}, R. 1986, \apj, 303, 233

\bibitem[{Skinner \& Brown(1994)}]{Skinner1994}
Skinner, S.~L., \& Brown, A. 1994, AJ, 107, 1461

\bibitem[{{Smith} {et~al.}(2003){Smith}, {Pestalozzi}, {G{\"u}del}, {Conway},
  \& {Benz}}]{Smith2003}
{Smith}, K., {Pestalozzi}, M., {G{\"u}del}, M., {Conway}, J., \& {Benz}, A.~O.
  2003, \aap, 406, 957

\bibitem[{Tasse {et~al.}(2013)Tasse, van~der Tol, van Zwieten, van Diepen, \&
  Bhatnagar}]{Tasse2013}
Tasse, C., van~der Tol, S., van Zwieten, J., van Diepen, G., \& Bhatnagar, S.
  2013, A{\&}A, 553, A105

\bibitem[{van Haarlem {et~al.}(2013)van Haarlem, Wise, Gunst, Heald, McKean,
  Hessels, de~Bruyn, Nijboer, Swinbank, Fallows, Brentjens, Nelles, Beck,
  Falcke, Fender, H{\"{o}}randel, Koopmans, Mann, Miley, R{\"{o}}ttgering,
  Stappers, Wijers, Zaroubi, van~den Akker, Alexov, Anderson, Anderson, van
  Ardenne, Arts, Asgekar, Avruch, Batejat, B{\"{a}}hren, Bell, Bell, van
  Bemmel, Bennema, Bentum, Bernardi, Best, B{\^{i}}rzan, Bonafede, Boonstra,
  Braun, Bregman, Breitling, van~de Brink, Broderick, Broekema, Brouw,
  Br{\"{u}}ggen, Butcher, van Cappellen, Ciardi, Coenen, Conway, Coolen,
  Corstanje, Damstra, Davies, Deller, Dettmar, van Diepen, Dijkstra, Donker,
  Doorduin, Dromer, Drost, van Duin, Eisl{\"{o}}ffel, van Enst, Ferrari,
  Frieswijk, Gankema, Garrett, de~Gasperin, Gerbers, de~Geus, Grie{\ss}meier,
  Grit, Gruppen, Hamaker, Hassall, Hoeft, Holties, Horneffer, van~der Horst,
  van Houwelingen, Huijgen, Iacobelli, Intema, Jackson, Jelic, de~Jong, Juette,
  Kant, Karastergiou, Koers, Kollen, Kondratiev, Kooistra, Koopman, Koster,
  Kuniyoshi, Kramer, Kuper, Lambropoulos, Law, van Leeuwen, Lemaitre, Loose,
  Maat, Macario, Markoff, Masters, McFadden, McKay-Bukowski, Meijering,
  Meulman, Mevius, Middelberg, Millenaar, Miller-Jones, Mohan, Mol, Morawietz,
  Morganti, Mulcahy, Mulder, Munk, Nieuwenhuis, van Nieuwpoort, Noordam,
  Norden, Noutsos, Offringa, Olofsson, Omar, Orr{\'{u}}, Overeem, Paas,
  Pandey-Pommier, Pandey, Pizzo, Polatidis, Rafferty, Rawlings, Reich,
  de~Reijer, Reitsma, Renting, Riemers, Rol, Romein, Roosjen, Ruiter, Scaife,
  van~der Schaaf, Scheers, Schellart, Schoenmakers, Schoonderbeek, Serylak,
  Shulevski, Sluman, Smirnov, Sobey, Spreeuw, Steinmetz, Sterks, Stiepel,
  Stuurwold, Tagger, Tang, Tasse, Thomas, Thoudam, Toribio, van~der Tol, Usov,
  van Veelen, van~der Veen, ter Veen, Verbiest, Vermeulen, Vermaas, Vocks,
  Vogt, de~Vos, van~der Wal, van Weeren, Weggemans, Weltevrede, White,
  Wijnholds, Wilhelmsson, Wucknitz, Yatawatta, Zarka, Zensus, \& van
  Zwieten}]{VanHaarlem2013}
van Haarlem, M.~P., Wise, M.~W., Gunst, A.~W., {et~al.} 2013, A{\&}A, 556, A2

\bibitem[{van Weeren {et~al.}(2016)van Weeren, Williams, Hardcastle, Shimwell,
  Rafferty, Sabater, Heald, Sridhar, Dijkema, Brunetti, Br{\"{u}}ggen,
  Andrade-Santos, Ogrean, R{\"{o}}ttgering, Dawson, Forman, de~Gasperin, Jones,
  Miley, Rudnick, Sarazin, Bonafede, Best, Bîrzan, Cassano, Chy{\.{z}}y,
  Croston, Ensslin, Ferrari, Hoeft, Horellou, Jarvis, Kraft, Mevius, Intema,
  Murray, Orr{\'{u}}, Pizzo, Simionescu, Stroe, van~der Tol, \&
  White}]{Weeren2016}
van Weeren, R.~J., Williams, W.~L., Hardcastle, M.~J., {et~al.} 2016, ApJS,
  223, 2

\end{thebibliography}

\end{document}